\documentclass[preprint, amsmath, amssymb, preprintnumbers, showpacs, showkeys, aps, prb]{revtex4-1}

\usepackage{graphicx}

\begin{document}

\title{Hard X-ray photoelectron spectroscopy on buried, off-stoichiometric
       Co$_x$Mn$_y$Ge$_z$ ($x:z=2:0.38$) Heusler thin films.}

\author{Siham Ouardi}
\affiliation{Max Planck Institute for Chemical Physics of Solids,
             D-01187 Dresden, Germany}
             
\author{Gerhard~H. Fecher}
\email{fecher@cpfs.mpg.de}
\affiliation{Max Planck Institute for Chemical Physics of Solids,
             D-01187 Dresden, Germany}

\author{Stanislav Chadov}
\affiliation{Max Planck Institute for Chemical Physics of Solids,
             D-01187 Dresden, Germany}

\author{Benjamin Balke}
\affiliation{Institut f\"ur Anorganische und Analytische Chemie,
Johannes Gutenberg - Universit{\"a}t, 55099 Mainz, Germany}

\author{Xenia Kozina}
\affiliation{Institut f\"ur Anorganische und Analytische Chemie,
Johannes Gutenberg - Universit{\"a}t, 55099 Mainz, Germany}

\author{Claudia Felser}
\affiliation{Max Planck Institute for Chemical Physics of Solids,
             D-01187 Dresden, Germany}

\author{Tomoyuki Taira}
\author{Masafumi Yamamoto}
\affiliation{Division of Electronics for Informatics,
             Hokkaido University, Sapporo 060-0814, Japan.}

\date{\today}

\begin{abstract}

Fully epitaxial magnetic tunnel junctions (MTJs) with off-stoichiometric 
Co$_2$-based Heusler alloy shows a intense dependency of the tunnel magnetoresistance 
(TMR) on the Mn composition, demonstrating giant TMR ratios of up to 1995\% at 4.2 K for ~\cite{LHT12}. 
This work reports on the electronic structure of non-stoichiometric
Co$_x$Mn$_y$Ge$_z$ thin films with a fixed Co/Ge ratio of $x:z=2:0.38$.
The electronic structure was investigated by high energy, hard X-ray photoelectron
spectroscopy combined with first-principles calculations. The high-resolution measurements of the
valence band of the non-stoichiometric Co$_x$Mn$_y$Ge$_z$
films close to the Fermi energy indicate a shift of the spectral
weight compared to bulk Co$_2$MnGe. This is in agreement with the
changes in the density of states predicted by the calculations.
Furthermore it is shown that the co-sputtering of Co$_2$MnGe together
with additional Mn is an appropriate technique to adjust the
stoichiometry of the Co$_x$Mn$_y$Ge$_z$ film composition. The
resulting changes of the electronic structure within the valence band
will allow to tune the magnetoresistive characteristics of
Co$_x$Mn$_y$Ge$_z$ based tunnel junctions as verified by the
calculations and photoemission experiments.

\end{abstract}



\maketitle

\section{Introduction}
\label{intro}

Nowadays, Co$_2$ based Heusler compounds are successfully used as ferromagnetic
electrodes providing a spin polarized electrical current in tunneling
magnetoresistive (TMR) junctions~\cite{IHM08,TeI09,ILT09}. To produce high
performance spintronic devices with Co$_2$ based Heusler thin films as
ferromagnetic electrodes, it is necessary to clarify the dependence of
the magnetoresistive characteristics on the stoichiometry of the
Heusler film electrode and its impact on the selection of materials for
TMR devices. However, very little is known about the electronic
structure of non-stoichiometric Heusler compounds.

Furthermore, partially substitution of one of the elements in 
the ternary Heusler $T_{2}T'M$ (where, $T,T'$ is transition metals and $M$ main group element)
 can be used to design new materials with predictable properties~\cite{FHK03,FKW05}, such as tuning the 
Fermi energy in to the middle of the minority band gap~\cite{FFe07}. 
The partially substitution of the elements in $T_{2}T'M$ leads to 
substitutional quaternary alloys of the type $T_{2}T'_{1-x}T''_xM$ or 
$T_{2}T'M_{1-x}M'_{x}$.
The quaternary alloy Co$_2$Cr$_{0.6}$Fe$_{0.4}$Al has attracted
emerged interest as potential material for magnetoelectronics, 
due to the large MR~\cite{BFJ03,BWF06}. The band structure
calculations confirmed the halfmetallic state of this compound 
in the ordered $L2_1$ structure~\cite{FHK03,FKW05,BFJ03}. 
Recently, several groups developed fully epitaxial magnetic tunnel junctions
(MTJs) based on Co$_2$Cr$_{0.6}$Fe$_{0.4}$Al as a lower electrode and a MgO 
tunnel barrier~\cite{ITO04,IOM06,IWJ08}.
However, a disorder of the structure results in a strongly reduced magnetic moment 
as well as a loss of the half-metallic character~\cite{EWF04a,EWF04b}.

Co$_2$MnZ (Z = Si, Ge, Sn) are representatives of Heusler 
compounds with a specific electronic band structure providing a pronounced
dependence of its transport properties on the orientation of the
electron spin~\cite{GMV83,KWS83b}. Structural and magnetic properties of
Co$_2$MnSi were reported for films and single
crystals~\cite{RRH02,WPK05} and the compound was suggested to be
suitable for magnetic tunnel junctions (MTJs)~\cite{IHM08,SHO06,TSO08}.
A high TMR ratio of 179\% at room temperature~(RT) and 683\% at 4.2~K 
was revealed by Ishikawa and coworkers~\cite{IHM08}. 
The Substitution of Fe by Mn in Co$_2$Fe$_x$Mn$_{1-x}$Si shows a 
dependency of half-metallicity and Gilbert damping constant on the 
composition $x$~\cite{KTO09,OKK10}.
Recently, anisotropic magnetoresistance (AMR) effect was systematically 
investigated in epitaxially grown Co$_2$Fe$_x$Mn$_{1-x}$Si Heusler films 
with respect to Fe composition $x$ and the annealing temperature~\cite{YSK12}.
It was shown that the AMR effect can be an indicator of half-metallicity.

An appropriate alternative to Co$_2$MnSi is the isovalent Co$_2$MnGe
with 29 valence electrons in the primitive cell. Recently, Co$_2$MnGe 
was investigated in greater detail by theoretical and experimental methods in order to prove the
half metallic ferromagnetic character, as well as their crystallographic structure, mechanical and transport
properties~\cite{OFB11}. This compound is of
interest for spintronic applications because it combines high Curie
temperature (905~K)~\cite{WEB71}, high magnetic moment
(5~$\mu_B$)~\cite{FKW06}, and coherent growth on top of
semiconductors~\cite{MRH08,MRH09}.
The non-stoichiometric Co$_2$MnGe was successfully used to produce
magnetic tunnel junctions~\cite{MIM06,IMM06,YaI10}. Hakamata and coworkers
fabricated epitaxial MTJs with a Co-rich Co$_2$MnGe film and reached
relatively low TMR ratios of 83\% at RT and 185\% at
4.2~K~\cite{HaI07}. Thereafter, Taira and coworkers  investigated the
influence of the annealing temperature on the performance of the
MTJs. They achieved a significant increase of the TMR up to 160\% at
RT (376\% at 4.2~K) by annealing at 500$^\circ$C~\cite{TAI09,TII09}.
This increase was explained in terms of the increase in the
interfacial spin polarization at the Fermi level associated with the
change in the spin-dependent interfacial density of states. On the
other hand, the effect of defects in Co$_2$Mn$_\beta$Ge$_\delta$ and Co$_2$Mn$_\alpha$Si films
on the TMR ratio  was extensively studied in Reference~\cite{YaI10}
where a strong dependence on the composition and Mn content was
observed. A high TMR ratio of 650\% at 4.2~K and 220\% at RT for Co$_2$Mn$_\beta$Ge$_\delta$ 
with $\beta$= 1.40 was found. The highest TMR ratios for Heusler based MTJs of 1135\% at
4.2 K and 236\% at RT was achieved for Mn-rich Co$_2$Mn$_\alpha$Si electrodes with $\alpha$ = 1.29.
It was suggested that detrimental Co$_{\rm Mn}$ antisites can be suppressed by preparing the films with a
Mn-rich composition.

The present work reports on the theoretical background and
experimental studies of the electronic structure of non-stoichiometric
Co$_2$Mn$_\beta$Ge$_\delta$. {\it ab-initio} electronic structure
calculations are used to verify the range of existence of the
half-metallic ferromagnetic behavior. High resolution photoelectron
spectroscopy of the core levels and valence states of buried
Co$_2$Mn$_\beta$Ge$_\delta$ films was used to explore the electronic
structure in comparison to the theoretical predictions. Hard X-ray
photoelectron spectroscopy (HAXPES) was previously shown to be a bulk
sensitive probe of the electronic structure~\cite{FBG08,OBG09,OGB09,OSF11,OFF12}.
The use of high-brilliance high-flux X-rays from the third-generation synchrotron 
radiation sources results the emission of electrons having high kinetic energies, in turn leading to a 
high probing depth because of the increased electron mean free 
path~\cite{DDB04}. Recently, several studies using high-resolution HAXPES have 
been realized. The electronic structure of solids like valence transitions in 
bulk systems~\cite{PST07} as well as multilayer systems~\cite{DDB04} and 
the valence band of buried thin films~\cite{FBG08} have been investigated. 
Newly, The use of polarized light in combination with HAXPES enables the analysis
of the symmetry of bulk electronic states~\cite{OFK11} as well as the
 magnetic properties of buried layers~\cite{KFS11}.
HAXPES was used in the present work to study buried
Co$_2$Mn$_\beta$Ge$_\delta$ films underneath MgO/AlO$_x$ capping
layers. The study of core level photoelectron spectra of the buried
Heusler thin films gives information on the electronic structure and
the chemical composition of these films.

\section{Experiment}
\label{sec:exp}

\begin{figure}[htb]
\centering
   \includegraphics[width=5cm]{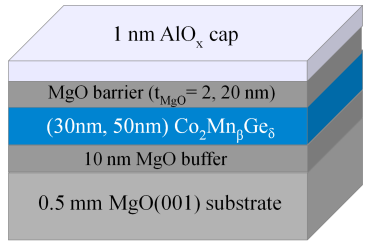}
   \caption{Schematic sample structure.}
\label{fig:sample}
\end{figure}

For the present study, special multi-component thin film arrangements
were produced that corresponded to a half of a magnetic tunnel junction as
used in TMR devices (see Figure~\ref{fig:sample}). In particular the
free electrode was modeled. The fabricated sample layer structure was
as follows: MgO(001) substrate (0.5~mm)~/ MgO buffer layer (10~nm)~/
Co$_2$Mn$_\beta$Ge$_\delta$ (30~nm or 50~nm)~/ MgO barrier ($t_{\rm MgO}$)~/ AlO$_x$
(1~nm) cap. $t_{\rm MgO}=2$~nm and 20~nm were chosen for the thickness of
the MgO layer. The topmost AlO$_x$ was used for protection of the
hygroscopic MgO layer during contact with air. The sample structure is
sketched in Figure~\ref{fig:sample}.

Each layer in the sample layer structure was successively deposited in
an ultrahigh vacuum chamber (base pressure: about $6 \times
10^{-8}$~Pa) through the combined use of magnetron sputtering for
Co$_2$MnGe and Al, and electron beam evaporation for MgO. The
fabrication procedure was the same as in previous
works~\cite{IMK06,IHM08}. The thin films were deposited at 300~K and
subsequently annealed in situ at $500^\circ$C for 15~min. Co-sputtering from
an additional Mn target was applied to vary the Mn content of the
Co$_2$Mn$_\beta$Ge$_\delta$ thin films. X-ray diffraction revealed
that the Co$_2$Mn$_\beta$Ge$_\delta$ films have a dominating $L2_1$
crystalline structure. The resulting Co$_2$MnGe layers used in the photoemission study were estimated
to be non-stoichiometric with the compositions of $2:1.03:0.38$,
$2:0.77:0.42$, and $2:0.67:0.38$ (see Reference~\cite{YaI10}). The MgO
barrier was deposited at 300~K. The AlO$_x$ cap was prepared by
exposing the sputter-deposited Al layer to an O$_2$ atmosphere of
$\approx 1\times 10^5$~Pa for 2 hours.

The precise occupation probability for each particular site of an assumed $L2_1$ structure
is unknown although the relative amounts of constituents are well-defined.
For this reason we will use the composition formula~\cite{YaI10}
derived from theoretical estimations~\cite{PCF04} of the defect formation
energies, using the permanent ratio of Co:Mn:Ge as $2:\beta:0.38$. 
Firstly, it is assumed that the Co-Mn-Ge alloy series 
basically does not contain vacancies. The Ge deficiency 
is compensated by Mn (Mn$_{\rm Ge}$ antisites). 
The lack of Mn in its native octahedral position 
is backfilled by Co$_{\rm Mn}$ antisites. Further, the lack
of Co in the tetrahedral sites is compensated by Mn$_{\rm Co}$ antisites.
This leads to the following compositions for $\delta=0.38$:
Co$_2$(Mn$_{1-x}$Co$_x$)(Ge$_{1-y}$Mn$_y$) for Mn deficit ($\beta\leq1.62$) and
(Co$_{1-x/2}$Mn$_{x/2}$)$_2$Mn(Ge$_{1-y}$Mn$_y$) for Mn excess ($\beta\geq1.62$). 
For $\delta=0.38$, the $x$ and $y$ antisite concentrations 
as function of the Mn content parameter $\beta$ 
are determined by the relations:

\begin{itemize} 
\item  $2+x:1-x+y:1-y=2:\beta:0.38\,\ (\beta\le 1.62)$;  
\item  $2-x:1+x+y:1-y=2:\beta:0.38\, \ (\beta\ge 1.62)$\,,
\end{itemize}

This corresponds to the experiments reported in Reference~\cite{YaI10}.

\begin{table*}[htbd]
	\centering
	\caption{ Selected site occupations of the primitive cell for the different
	          $\beta$ values of Co$_2$Mn$_\beta$Ge$_{0.38}$. \newline
	          The ratio of the Co:Ge content is fixed at $2:0.38$ and all sites are completely
	          occupied. Ge is placed exclusively on the 4b position.
	          The site occupation for the compound with Co:Ge~$=2:0.42$
	          and the original $L2_1$ structure are included for completeness.}

     \begin{tabular}{lccccccc}
     Composition                       & $\beta$ & Co (8c) & Co (4a) & Mn (4a) & Mn (8c) & Ge (4b)  & Mn (4b) \\
     \hline
     Co$_{3}$Mn$_{0.43}$Ge$_{0.57}$    & 0.28667 & 100\%   & 100\%   &  0      &  0      &  57\%    & 43\%  \\    
     Co$_{2.62}$Mn$_{0.88}$Ge$_{0.49}$ & 0.67    & 100\%   &  62\%   &  38\%   &  0      &  50\%    & 50\%  \\
     Co$_{2.5}$Mn$_{0.97}$Ge$_{0.53}$  & 0.77, $\delta=0.42$ & 100\%   &  51\%   &  49\%   &  0      &  52\%    & 48\%  \\ 
     Co$_{2.37}$Mn$_{1.18}$Ge$_{0.45}$ & 1       & 100\%   &  37\%   &  63\%   &  0      &  45\%    & 55\%  \\    
     Co$_{2.35}$Mn$_{1.21}$Ge$_{0.44}$ & 1.03    & 100\%   &  35\%   &  65\%   &  0      &  45\%    & 55\%  \\
     Co$_{2}$Mn$_{1.62}$Ge$_{0.38}$    & 1.62    & 100\%   &   0     & 100\%   &  0      &  38\%    & 62\%  \\
     Co$_{1.82}$Mn$_{1.82}$Ge$_{0.35}$ & 2       &  91\%   &   0     & 100\%   &  9\%    &  35\%    & 65\%  \\
     Co$_{1.68}$Mn$_{2}$Ge$_{0.32}$    & 2.38    &  84\%   &   0     & 100\%   & 16\%    &  32\%    & 68\%  \\     
     \hline
     Co$_2$MnGe                        & $L2_1$  & 100\%   &         & 100\%   &          & 100\%   & \\     
	   \end{tabular}
\label{tab:sym}	
\end{table*}

In the regular $L2_1$ structure of the X$_2$YZ Heusler compounds, the
atoms occupy the Wyckoff positions 8a (X$_2$) ($T_d$), 4a (Y) ($O_h$),
and 4b (Z) ($O_h$) of the primitive cell with $F\:m\overline{3}m$
symmetry (space group 225). Table~\ref{tab:sym} summarizes the site
occupations of the primitive cell for the different $\beta$ values.
$\beta=0.28\overline{6}$ corresponds to a $DO_3$ structure according to
Co$_3$(Mn$_{0.43}$Ge$_{0.57}$). It should be noted that for
compositions with $\rm{Mn:Co}>2$ ($\beta>1.62$) a $Y$-type structure with $F\:\overline{4}3m$
symmetry (space group 216) can be expected where the 8c position of
the $L2_1$ structure is split into 4c and 4d such that all sites
(4a-4d) have $T_d$ symmetry. In that case Mn may occupy 4a and 4c
whereas Co and Ge are distributed on 4b and 4d. Further, it is known
that the binary alloys Co$_{0.5}$Mn$_{0.5}$~\cite{CT95} or Co$_{0.87}$Ge$_{0.13}$~\cite{LD63} 
crystallize in $fcc$ lattices with random site occupation and similar lattice parameters. 
Therefore, other types of disorder or site occupations cannot be excluded. Due to the similar
scattering amplitudes of the constituents, however, X-ray diffraction
is not unambiguous in the structure determination.

\section{Calculations of the electronic structure}

To understand the electronic structure of the non stoichiometric compounds,
first-principles calculations were performed using the KKR
(Korringa-Kohn-Rostoker) Green's function method as implemented in the
Munich SPR-KKR (spin-polarized relativistic) package~\cite{HM96,EBE99}. The
chemical disorder is treated by the coherent potential approximation
(CPA)~\cite{BG72,WH85}. Exchange and correlation are treated within the
local spin density approximation (LSDA) using  the parametrization of
Vosko, Wilk and Nussair~\cite{SLM80}. The full symmetry potential
method was used in order to account for the non spherical character of
the atomic potentials in solids that is rather significant in Heusler
compounds. The $k$-integration mesh was set to a size of
$(22\times22\times22)$ during the selfconsistent cycles resulting in
328 points in the irreducible wedge of the Brillouin zone. The
expansion in spherical harmonics up to $l_{\rm max}=3$ ($f$-electrons)
was found to be numerically sufficient. The calculations were performed
for a disordered $L2_1$ structure to account for the composition dependence. 
The site occupations for the different compositions are explained in detail in Section~\ref{sec:exp}.

\begin{figure}[htbd]
\centering
 \includegraphics[width=7cm]{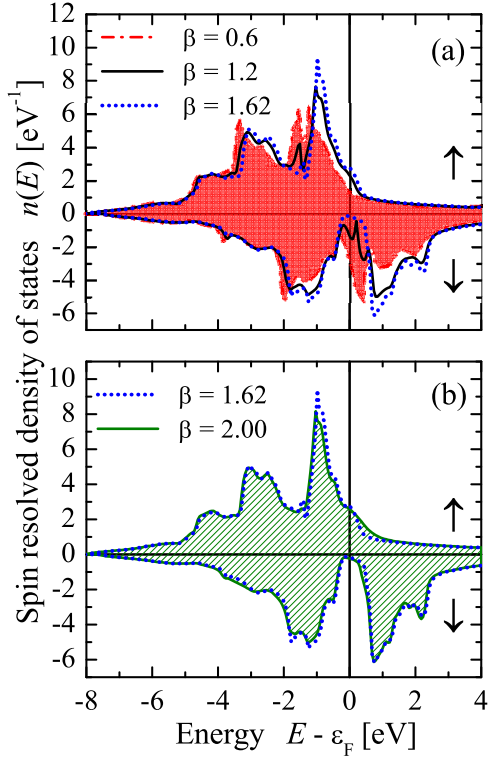}
  \caption{(color online) Spin-resolved DOS calculated 
  for (a) Mn-deficient  ($\beta \leq 1.62$) series, (b) Mn-excessive
  ($\beta = 1.62, 2.0$) series. Minority states ($\downarrow$) 
  are shown on a negative scale.}
\label{FIG:DOS_SPIN-RESOLVED}
\end{figure}

\begin{figure}[htbd]
\centering
  \includegraphics[width=7cm]{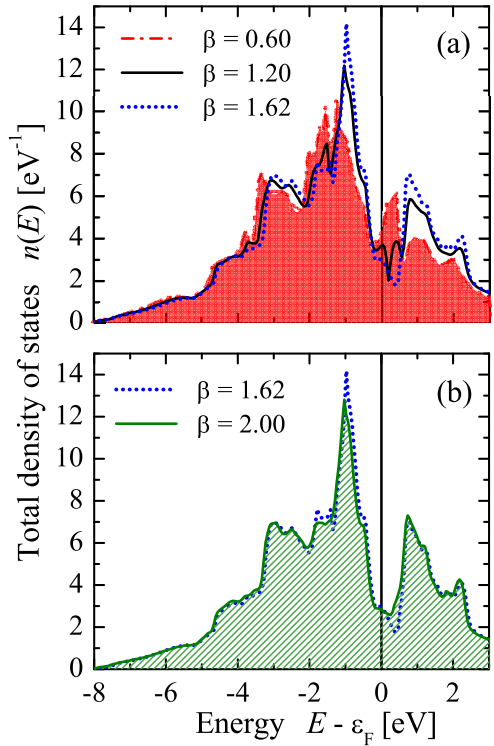}
  \caption{(color online) Total DOS (sum of both spin channels) calculated 
  for (a) Mn-defficient  ($\beta \leq 1.62$) (b) Mn-excess
  ($\beta = 1.62, 2.0$).}
\label{FIG:DOS_SPIN-SUMMED}
\end{figure}

\begin{figure}[htbd]
\centering
  \includegraphics[width=8cm]{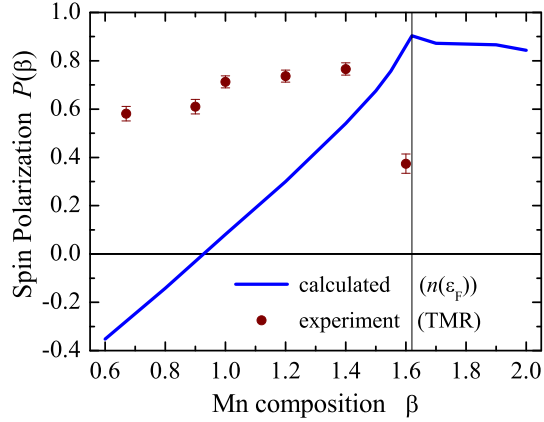}
   \caption{Spin-polarization $P$ at the Fermi energy 
calculated as a function of Mn composition $\beta$.\newline
Compared are the values from ab-initio calculations to
experimental values calculated from the $TMR$ measured at 4.2~K
(see Ref.:~\cite{YaI10}) using $P_{TMR}=TMR / (TMR+2)$.
The vertical line assigns $\beta=1.62$.}
\label{FIG:SPINPOL}
\end{figure}

\begin{figure}[htbd]
\centering
  \includegraphics[width=6cm]{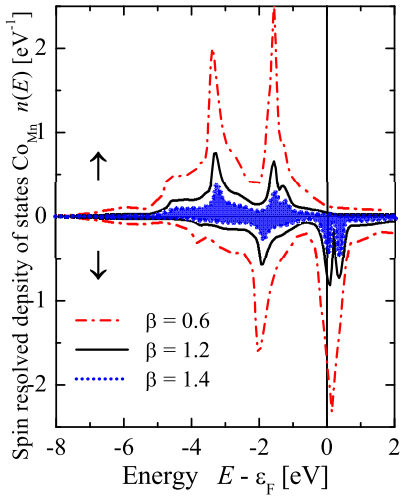}
  \caption{(color online) Spin-resolved DOS of Co$_{\rm Mn}$ (Co antisites)
  for Mn-deficient regime ($\beta < 1.62$). Minority states ($\downarrow$) are shown on a negative scale.}
\label{FIG:DOS_CO-IMPURITY}
\end{figure}

The results of the electronic structure calculations are discussed in
the following. It is obvious from the comparison of the density of
state (Figures~\ref{FIG:DOS_SPIN-RESOLVED}(a) and (b)) that  the most
significant factor influencing the electronic properties is the
formation of Co$_{\rm Mn}$-antisites in the Mn-poor regime
($\beta<1.62$). In full analogy to Co-Mn-Si alloy
systems~\cite{PCF04,CT08}, Co$_{\rm Mn}$-antisites introduce
impurity-like states (shown separately in
Figure~\ref{FIG:DOS_CO-IMPURITY}) inside of the half-metallic band gap
that seriously destroy the spin-polarization (see
Figure~\ref{FIG:SPINPOL}: $\beta<1.62$). In general, the spin polarisation 
at the Fermi energy is defined by:

\begin{equation}
	P(\epsilon_{\rm F}) = \frac{v^x_\uparrow n_\uparrow(\epsilon_{\rm F}) - v^x_\downarrow n_\downarrow(\epsilon_{\rm F})}
	                          {v^x_\uparrow n_\uparrow(\epsilon_{\rm F}) + v^x_\downarrow n_\downarrow(\epsilon_{\rm F})},
\end{equation}

where $v_\uparrow$ and $v_\downarrow$ are the Fermi velocities for majority 
and minority electrons, respectively. In photoemission one has $x=0$ such 
that the spin polarisation is purely given by the majority ($n_\uparrow$) 
and minority ($n_\downarrow$) densities. The situation is, however, 
different for the TMR wherefore the values can not be compared
directly but only their trend. It is worthwhile to note that in 
halfmetallic ferromagnets one has $P(\epsilon_{\rm F})=1$ independent
of the method.

\begin{figure}[htbd]
\centering
  \includegraphics[width=9cm]{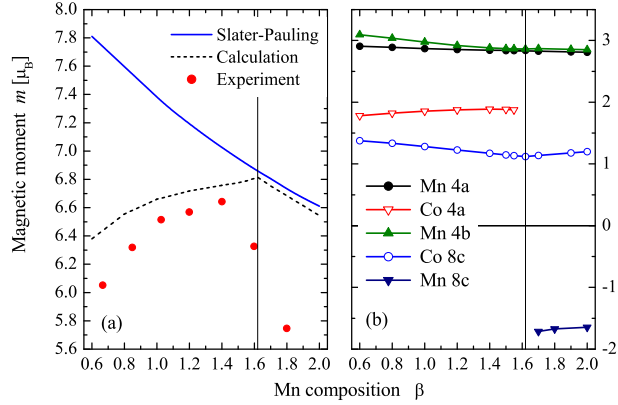}
  \caption{(color online) (a) Total magnetic moments calculated as a
  function of Mn composition $\beta$ (experimental data from~\cite{YaI10});
  (b) Calculated partial magnetic moments.
  (See Table~\ref{tab:sym} for the notation of the Wyckoff positions  4a, 4b, 8c
  and their site occupations;
  vertical lines are at $\beta=1.62$.)}
\label{FIG:MOMENTS}
\end{figure}

The total magnetic moment in
the Mn-deficient alloy strongly deviates from the Slater-Pauling rule
(Figure~\ref{FIG:MOMENTS}(a): $\beta<1.62$). This reasonably agrees
with the measured TMR ratio and magnetic moments~\cite{YaI10}.

In contrast, the excess of Mn (which result in a formation of
Mn$_{\rm Co}$ antisites) does not influence the half-metallic gap
noticeable (Figure~\ref{FIG:DOS_SPIN-RESOLVED}(b)). The
system remains nearly half-metallic. This restores the
Slater-Pauling behavior of the total magnetic moment
(Figure~\ref{FIG:MOMENTS}(a): $\beta>1.62$). 
The slightly reduced value of the spin-polarization for $\beta\ge 1.62$
is mainly caused by Mn$_{\rm Ge}$ antisites
permanently present in the whole series. This is also
reflected in a slightly reduced value of the total magnetic moment
compared to the Slater-Pauling rule.

As it follows from the calculated atomic-resolved magnetic moments
shown in Figure~\ref{FIG:MOMENTS}(b), the total magnetization is
mainly controlled by the amount of Co$_{\rm Mn}$ and Mn$_{\rm Co}$
antisites. For the Mn-deficient system the strong moment at the Mn
atom (about $3~\mu_{\rm B}$) in the octahedral position is 
accompanied by the lower moment of the Co$_{\rm Mn}$ antisite  (about
$1.9~\mu_{\rm B}$). This is the reason for the violation of the
Slater-Pauling  rule. On the other hand, all possible Mn permutations
always result into half-metallic contributions. For the Mn-excessive
alloys, formation of the Mn$_{\rm Co}$ antisites at tetrahedral
positions exhibit a negative moment  of about $-1.7~\mu_{\rm B}$. The
ferrimagnetic coupling between the regular Mn atom and the Mn$_{\rm
Co}$ antisite appears because they are nearest neighbors. Obviously,
the magnetic interaction between neighboring Co atoms is smaller when
part of them occupies Co$_{\rm Mn}$ antisites due to the smaller
extend of the $d$-wave functions.

From the comparison of the total (spin-integrated) density of states
(see Figure~\ref{FIG:DOS_SPIN-SUMMED}) and the photoelectron spectra
(Figure~\ref{fig:ef8kev}), it was found that the Mn $e_g$ states
decrease with decreasing Mn content. For the Mn-poor alloys the
minimum of the density of state at the Fermi energy is destroyed by
the appearance of Co$_{\rm Mn}$ antisites. With increasing the Mn
content the model system behaves as a half-metal following the
Slater-Pauling scenario. The rapid decrease of the measured
magnetization~\cite{YaI10} above $\beta=1.4$ (see
Figure~\ref{FIG:MOMENTS}(a)) may be attributed to difference types of
disorder including possible structural distortions in thin films.

\section{Hard X-ray photoelectron spectroscopy}
\label{sec:haxpes}

The HAXPES experiments were performed at the undulator beamlines
BL15XU and BL47XU of SPring-8. At BL15XU, the photon energy was fixed
at 5.9469~keV using a Si(111) double crystal monochromator (DCM) and a
the 333 reflection of a Si channel-cut post monochromator. At
BL47XU, the photon energy was fixed at 7.9382~keV using a Si(111) DCM
and a Si(444) channel-cut post monochromator. The photo-emitted
electrons were analyzed for their kinetic energy and detected by a
hemispherical analyzer (VG Scienta R4000). The overall energy
resolution (monochromator plus analyzer) was set to 250~meV, as
verified by spectra of the Au valence band at the Fermi energy
($\epsilon_{\rm F}$). Additionally, spectra close to the Fermi
energy were taken with a total energy resolution of 150~meV. The angle
between the electron spectrometer and photon propagation is fixed at
$90^\circ$ in all experiments. A near normal emission
($\theta=2^\circ$) detection angle was used resulting in an angle of
photon incidence of $\alpha=88^\circ$. The measurements were taken at
sample temperatures of 20~K and 300~K.

\subsection{Core level spectroscopy}

Hard X-ray core level spectra were measured
on Co$_2$Mn$_\beta$Ge$_\delta$ films with different compositions to investigate
the influence of the off-stoichiometry on the spin orbit splitting as
well as the exchange interaction of the unpaired valence electrons with
the core holes.

\begin{figure}[htb]
\centering
   \includegraphics[width=8cm]{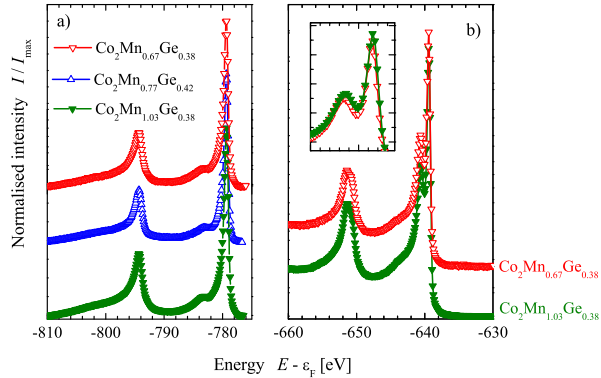}
   \caption{HAXPES spectra of the Co $2p$ (a) and Mn $2p$ (b) core level of Co$_2$Mn$_\beta$Ge$_\delta$ films.
     The spectra are taken at room temperature. The excitation energy was set to $h\nu=7.9382$~keV.
     The inset in (b) shows the details of the Mn $2p_{3/2}$ peak on an enlarged scale. }
\label{fig:CL}
\end{figure}

Figures~\ref{fig:CL}(a) and~(b) compare the hard X-ray Co $2p$ and Mn
$2p$ core level spectra of the non-stoichiometric
Co$_2$Mn$_\beta$Ge$_\delta$ films covered with 2~nm MgO. The spectra
were taken at room temperature with an excitation energy of about
8~keV. Concentrating on the Co $2p$ region, Figure~\ref{fig:CL}(a)
illustrates a spin-orbit splitting between the Co $2p_{3/2}$ and
$2p_{1/2}$ states of about 14.86~eV. There is no remarkable change of
the splitting or line width detected for the different
non-stoichiometric films. All spectra exhibit a satellite at about
4~eV below the center of the $2p_{3/2}$ main line. Such satellites
were also observed in the previous work~\cite{OBG09,OGB09}. The
satellite emerges from the interaction of the $2p$ core hole with the
partially filled $d$ valence bands.  To clarify the detailed origin of
the satellite structure, theoretical calculations are required that 
treat the details of the exchange interaction between the Co $3d$ electron 
occupation (in other words, the band structure) and $2p$ core hole states.
At the $2p_{1/2}$ state it is only seen weakly as a tail at
the low energy side due to shorter lifetime of the core hole compared
with the $2p_{3/2}$ state.

Figure~\ref{fig:CL}(b) shows the Mn $2p$ photoelectron spectra for the
buried Heusler thin films Co$_2$Mn$_{1.03}$Ge$_{0.38}$ and
Co$_2$Mn$_{0.67}$Ge$_{0.38}$. The peak intensity of the spin-orbit split $2p$ states
of the Mn-rich film Co$_2$Mn$_{1.03}$Ge$_{0.38}$ were slightly higher
than those with less Mn. This is in agreement with
the estimated composition of the films. In analogy to the Co $2p$
spectra, the Mn $2p$ core level shows the spin-orbit splitting of about
12.1~eV. Different to the Co $2p$ spectra, the multiplet
splitting of the Mn $2p_{3/2}$ state was clearly revealed (see inset in
Figure~\ref{fig:CL}(b)). It was shown theoretically and experimentally
for the case of Mn that the Coulomb interaction of the $2p$ core hole
and the $3d$ valence electrons leads to the splitting of the Mn
$2p_{3/2}$ level into several main sublevels~\cite{FSH70} caused by
existence of more than one possible final ionic state during ejection
of an electron from the $p$ shell. The effect of the exchange
interaction between $2p$ core hole and the partially filled $d$-bands
is less pronounced at the $2p_{1/2}$ state but still clearly
detectable. The small influence of the composition on the Mn $2p$ 
derived multiplet is in agreement with the electronic structure 
calculations where the localized Mn $d$ bands and the magnetic moment 
at Mn are stable against variation of the composition.
This finding underlines the localized character of the Mn valence $d$ electrons.

\begin{figure}[htbd]
\centering
   \includegraphics[width=9cm]{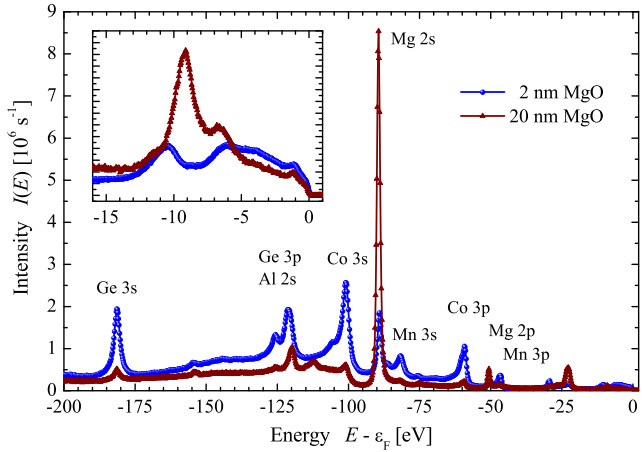}
   \caption{HAXPES spectra of the shallow core states of buried Co$_2$Mn$_{0.77}$Ge$_{0.42}$ thin films.
       The spectra with different thickness of the MgO layer (2~nm and 20~nm) are compared.
       The excitation energy was set to 7.9392~keV. The small peak at about -25~eV is due to the O $2s$ excitation.
       The inset compares the valence band spectra
       for different MgO thickness.}
\label{fig:vbMgO}
\end{figure}

Figure~\ref{fig:vbMgO} compares the hard X-ray photoelectron spectra of
the shallow core levels with low binding energy of the 50~nm thick
Co$_2$Mn$_{0.77}$Ge$_{0.42}$ layer buried underneath MgO/AlO$_x$ with
different thicknesses of the MgO interlayer. The intensity ratios of
the core level $I^{\rm{Mg}}_{2s}/I^{\rm{Ge}}_{3s}$ of the films covered
with 2~nm or 20~nm thick MgO are 0.95 or 16.45, respectively. This
strong increase is caused by the increase of the emission from the
thick MgO layer. The inset of Figure~\ref{fig:vbMgO} shows the valence
band of the covered thin films. The high intensity at -1.3~eV below
$\epsilon_{\rm F}$ is clearly resolved and agrees well with the bulk spectra
(see also: Figures~\ref{fig:vb8kev} and~\ref{fig:ef8kev}). The low
lying $s$ band below -8~eV is only seen for the film with the 2~nm
thick MgO layer. This part, in addition to the lower parts of the $p$
and $d$ bands, is covered by the emission from the Mg $3s$ and O
$2p$ states when the MgO film is 20~nm thick. This has been already reported
for other Heusler compounds~\cite{FBG08} and makes clear that the
valence band spectra close to the Fermi energy are not influenced by
the insulating layers above.

\subsection{Valence band spectroscopy}
\label{sec:vb}

\begin{figure}[htbd]
\centering
   \includegraphics[width=6cm]{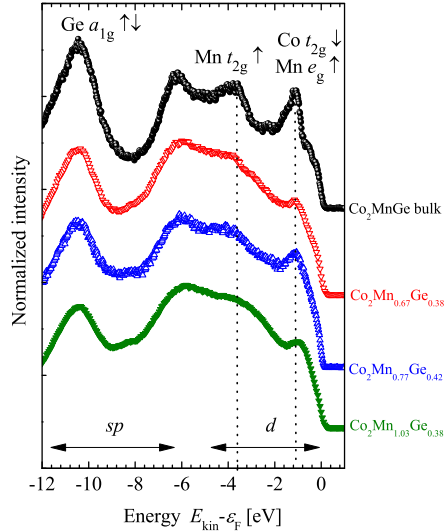}
   \caption{Valence band spectra of the
            Co$_2$Mn$_\beta$Ge$_\delta$ films and Co$_2$MnGe bulk sample.
            The excitation energy was set to 7.9392~keV.}
\label{fig:vb8kev}
\end{figure}

Figure~\ref{fig:vb8kev} compares the valence band spectra of the
non-stoichiometric Co$_2$Mn$_\beta$Ge$_\delta$ films to stoichiometric
bulk Co$_2$MnGe. The main features of the valence band of the thin
films agree well with that of the bulk sample. The intense, broad band
at about -10.5~eV is observed for the buried films as well as for the bulk
sample. This band is related to Ge $a_{1g}$ ($s$) states and is
separated from the remaining valence bands by the $sp$ hybridization
gap at around -8.5~eV. This low lying band gap is characteristic for
well ordered Heusler compounds with $L2_1$ structure. This
hybridization gap is well pronounced for the bulk sample, confirming
the ordered structure. Additional features within the $sp$
hybridization gap are seen in the spectra of the buried films.
Partially, this intensity arises from the O $2p$ states of the oxide
overlayers and thus a direct comparison to the {\it clean} bulk
material is not straightforward. However, the difference in the
spectral shapes between the films with altered composition one has the
same overlayer and differences in the spectra can be considered as an
evidence for the disorder rising up when changing the composition. The
spectral shape above about -6~eV are not influenced by the MgO and
AlO$_x$ layers as already mentioned above. Overall, the spectra of the
Co$_2$Mn$_\beta$Ge$_\delta$ films show broadened structures compared
to the spectrum of bulk Co$_2$MnGe. This broadening of the bands is
explained by the disturbed translational periodicity due to the random
site occupation or periodically unoccupied sites in the
non-stoichiometric Co$_2$Mn$_\beta$Ge$_\delta$ films. The majority Mn
$t_{2g}$ states at about -3.9~eV, that are responsible for the
localized magnetic moment at the Mn atoms, are much less pronounced in
the non-stoichiometric films compared to stoichiometric Co$_2$MnGe.
All the states far below $\epsilon_{\rm F}$ influence mainly the
magnetic but not directly the transport properties.

\begin{figure}[htbd]
\centering
   \includegraphics[width=5cm]{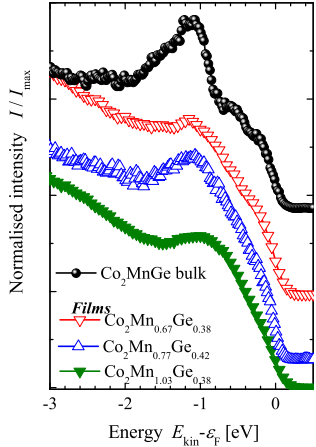}
   \caption{Valence band spectra close to $\epsilon_{\rm F}$.
            The spectra are taken for Co$_2$MnGe bulk as well as thin Co$_2$Mn$_{1.03}$Ge$_{0.38}$,
            Co$_2$Mn$_{0.77}$Ge$_{0.42}$, and Co$_2$Mn$_{0.67}$Ge$_{0.38}$ films.
            The excitation energy was set to 7.9392~keV.}
\label{fig:ef8kev}
\end{figure}

Most interesting with respect to the transport properties, in 
particular the use of the films in TMR devices, is the energy range
close to the Fermi energy. The maximum in the density of the $d$ states at
-1.1~eV is clearly resolved in the spectra from all films and the 
intensity varies when Mn content is changed (see
Figure~\ref{fig:ef8kev}). This maximum in the intensity arises from Co
$t_{2g}$ minority and Mn $e_g$ majority states. Obviously, those states
appear much stronger and sharper in the stoichiometric bulk sample
pointing to their higher localization in the defect free sample. The
intensity directly at $\epsilon_{\rm F}$ is dominated by $t_{2g}$ majority
states that are mainly located in the Co$_2$ planes of the compound.
The increase of the Mn content enhances the intensity of the highest
majority $d$-band that is mainly responsible for the electronic
transport properties. Such composition dependent behavior of the
density of states close to $\epsilon_{\rm F}$ provides the possibility to
tune the magnetoresistive characteristics of tunnel junctions by
changing the composition of the Co$_2$Mn$_\beta$Ge$_\delta$ electrode,
as was previously proposed in Reference~\cite{YaI10}.

\begin{figure}[htbd]
\centering
\includegraphics[width=8cm]{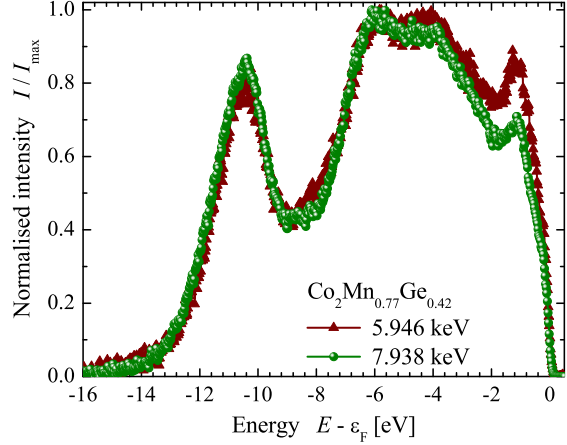}
   \caption{ Comparison of the valence band
      spectra of the Co$_2$Mn$_{0.77}$Ge$_{0.42}$ film excited by different
      photon energies (5.9469~keV and 7.9382~keV). }
\label{fig:vb_comp}
\end{figure}

Figure~\ref{fig:vb_comp} compares the valence band spectra on the
buried Co$_2$Mn$_{0.77}$Ge$_{0.42}$ film taken with different photon
energies of about 5.946~keV and 7.939~keV. The differences in the
spectra are caused by the different weights of the partial 
cross-sections for $s$, $p$, and $d$ electrons at different excitation
energies~\cite{FKW06}. With increasing energies the cross-sections for
$d$ electron excitation decrease faster than those for $s$ or $p$
electrons. Therefore, the contribution from $d$ electrons is more
pronounced at lower energy photo-excitation. The observed shift of
spectral weight towards the Fermi energy in the non-stoichiometric
films compared to bulk Co$_2$MnGe is in agreement with the changes in the
density of states predicted by the calculations.

\begin{figure}[htbd]
\centering
\includegraphics[width=8cm]{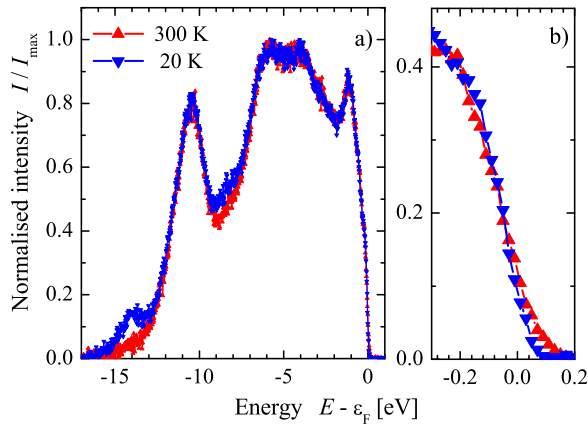}
   \caption{Valence band spectra of Co$_2$Mn$_{0.77}$Ge$_{0.42}$ film.
      (a)~compares the spectra measured at different temperatures ($h\nu=5.9469$~keV).
      (b)~shows the energy range close to the Fermi energy on an enlarged scale.
      }
\label{fig:vb_temp}
\end{figure}

Furthermore, the valence band spectra of the\\
Co$_2$Mn$_{0.77}$Ge$_{0.42}$ thin films taken at room temperature and
at 20~K are compared in Figure~\ref{fig:vb_temp}. No general changes in
the electronic structure are observed in the spectra while changing
temperature. At $T=20$~K a small, additional peak appears that is 
caused by adsorbed carbon oxides on the surface of the samples
while cooling. This observation provides evidence for the requirement
on ultra high vacuum for HAXPES measurement even though this technique is bulk
sensitive. Indeed, a massive contamination of the surface in the order
of several layers will influence the HAXPES spectra.

Figure~\ref{fig:vb_temp}(b) shows a broadening of the spectra close to
the Fermi energy. The broad spectral feature in the vicinity of the
Fermi energy at 300~K compared to 20~K is caused by the temperature
dependence of the Fermi-Dirac distribution. This observation is in
agreement with reports showing that Co$_2$MnSi films do not exhibit
temperature effects in photoelectron spectroscopy~\cite{MKM09}.

\section{Summary and conclusions}

In summary, the electronic structure of off-stoichiometric
Co$_2$Mn$_\beta$Ge$_\delta$ alloys was investigated by first-principles calculations
and bulk sensitive photoelectron spectroscopy on thin
films. The electronic structure calculations demonstrate that the
magnetic moments of Co and Mn atoms (including the antisites) are
nearly independent of the stoichiometry. This finding is in agreement
with the core level photoelectron spectra where no differences in the
multiplet splitting of the Mn $2p$ states was observed for different
compositions.

The formation of Co$_{\rm Mn}$ antisites was found to be the most
destructive depolarizing mechanism. Corresponding results for the
Mn-deficient alloy are in reasonable agreement with earlier
measurements of  TMR and magnetization. This finding also agrees
with the valence band spectra which exhibit a noticeable smearing of
the Mn $t_{2g}$ and $e_g$ states for the Mn-deficient alloys. On the
other hand, the relative decrease of the measured  magnetization,
especially the significant violation of the Slater-Pauling rule in the
Mn-excessive regime, indicate the presence of additional disorder
mechanisms different from the assumed types of antisites disorder. 
The valence band spectra of the buried Co$_2$Mn$_\beta$Ge$_\delta$
thin films exhibit still some major differences compared to
stoichiometric bulk Co$_2$MnGe. However, the electronic structure of
the non-stoichiometric films was found to reveal the major
characteristic features of the bulk analogue. The observed shift of
spectral weight towards the Fermi energy in the non-stoichiometric
films compared to bulk Co$_2$MnGe is in agreement with the changes in
the density of states predicted by the calculations. 
Temperature effects on the states at the Fermi energy that go beyond
the broadening of the Fermi-Dirac distributions are not observed. This
is, indeed, expected from the high Curie temperature of the compound.

Co-sputtering of Co$_2$MnGe together with additional Mn is shown to be
a well suited technique to tune the stoichiometry of the
Co$_2$Mn$_\beta$Ge$_\delta$ film composition that provides the
possibility to change the electronic structure within the valence band
and as a consequence the magnetoresistive characteristics of
Co$_2$Mn$_\beta$Ge$_\delta$ based tunnel junctions.

\bigskip
\begin{acknowledgements}
The authors thank E. Ikenaga (BL47XU) and S. Ueda (BL15XU) for help with the experiments.
This work was funded by the {\it Deutsche Forschungs Gemeinschaft DFG}
(TP~1.2-A and 1.3-A of the Research Unit ASPIMATT) and the 
{\it Japan Science and Technology Agency JST} (DFG-JST project: FE633/6-1).
The work at Hokkaido University was partly supported by a Grant-in-Aid for
Scientific Research (A) (Grant No. 20246054) from the MEXT, Japan, and by the 
Strategic International Cooperative Program of JST. The synchrotron radiation 
HAXPES measurements were performed at BL47XU with the approval of the Japan 
Synchrotron Radiation Research Institute (JASRI) (Long-term Proposal 
2008B0017, 2009A0017) and at BL15XU with the approval of NIMS (Nanonet Support
Proposal 2008B4903). The HAXPES experiment at BL15XU was partially
supported by the Nanotechnology Network Project MEXT (Japan).
\end{acknowledgements}

%

\end{document}